\documentclass[unnumsec,webpdf,contemporary,large,numbered]{oup-authoring-template}% uncomment this line to get the numbered reference output
\graphicspath{{figs/}}

\usepackage{hyperref}
\usepackage{booktabs}
\usepackage{graphicx}

\theoremstyle{thmstyleone}%
\theoremstyle{thmstyletwo}%
\theoremstyle{thmstylethree}%

\begin{document}

\journaltitle{Title of Journal}
\DOI{DOI added during production}
\copyrightyear{YEAR}
\pubyear{YEAR}
\vol{XX}
\issue{x}
\access{Published: Date added during production}
\appnotes{Paper}

\firstpage{1}

%\subtitle{Subject Section}

\title[GenomeHarness]{GenomeHarness: Harnessing AI Agents for Reliable Adaptation of Genome Language Models}

\author[1]{Weicai Long}  % $\dagger$
\author[1]{Yusen Hou}
\author[1]{Houcheng Su}
\author[1]{Junning Feng}
\author[$\ast$]{Yanlin Zhang.}

\address[1]{Data Science and Analytics Thrust, The Hong Kong University of Science and Technology (Guangzhou), Guangzhou, China}

% \corresp[$\dagger$]{The authors contribute equally to this work.}

\corresp[$\ast$]{Corresponding author: \href{yanlinzhang@hkust-gz.edu.cn}{yanlinzhang@hkust-gz.edu.cn}}

\abstract{
Pretrained genome language models provide reusable representations for DNA sequence analysis, but turning them into reliable downstream predictors remains non-trivial. Their practical performance depends strongly on fine-tuning recipes, and default recipes reported in prior studies may be suboptimal for new tasks or model backbones, making weak downstream results difficult to interpret. Beyond recipe choice, users must navigate a complex adaptation workflow such as data preparation, environment setup, compute management, failure diagnosis, and comparing stochastic runs without test leakage. These requirements place a substantial operational burden on many intended users, whose expertise is often centered on biological questions and interpretation rather than machine-learning engineering. Reliable use of genome language models therefore requires more than conventional AutoML-style tuning: it requires a systematic, budget-aware, and auditable procedure that lowers the barrier to downstream adaptation.
We present GenomeHarness, an agentic harness for adapting genome language models through controlled search over fine-tuning recipes. GenomeHarness combines an AI agent for proposing and repairing recipe edits, a harness for protocol-constrained execution, resource management, and test isolation, and a Monte Carlo tree search controller for allocating search effort across recipe lineages. We evaluate GenomeHarness on DNABERT2 and NTv2-100M-Multi across the NT Benchmark and Genomic Benchmarks. Final evaluation is performed using three random seeds after recipe freezing. Across 52 model-task settings, GenomeHarness improves mean test MCC in 47 settings, including 24 of 26 DNABERT2 settings and 23 of 26 NTv2-100M-Multi settings. The gains are especially pronounced on Genomic Benchmarks and on tasks where the root recipe is unstable or poorly matched, such as human\_ocr\_ensembl task. Search traces further show that GenomeHarness progressively identifies stronger recipes, turning downstream adaptation into a controlled and auditable workflow rather than a manual tuning process.
The code is Available at https://github.com/ai4nucleome/GenomeHarness.\\
} 

% \textbf{Supplementary Information:} yanlinzhang@hkust-gz.edu.cn\\

% \keywords{Topologically Associating Domain, Paired Boundary, TAD caller, TAD Annotation, Hi-C contact map}

% \keywords[Abbreviations]{abbreviation1, abbreviation2, abbreviation3, abbreviation4}

% \otherabstract[Graphical Abstract]{\colorbox{black!20}{\hbox to 0.97\textwidth{\vbox to 50pt{}}}}

% \boxedtext{Key Messages}{
% \begin{itemize}
% \item Key boxed text here.
% \item Key boxed text here.
% \item Key boxed text here.
% \end{itemize}}

\maketitle

%\begin{epigraph}
%Epigraph text. Ximporem qui reperov idempedit modio. Bisto imagnatem quae aceptis
%nobitae quid eum rae adignis quias-sit vellacc uptatur sunt quis rentis eaquasit alia deliquam
%rec-to consed unt. Empor sum ratur ressimusdae. Nam fugiae.
%\source{Epigraph source}
%\end{epigraph}

\section{Introduction}

\begin{figure*}[ht]
    \centering
    \includegraphics[width=0.8\textwidth]{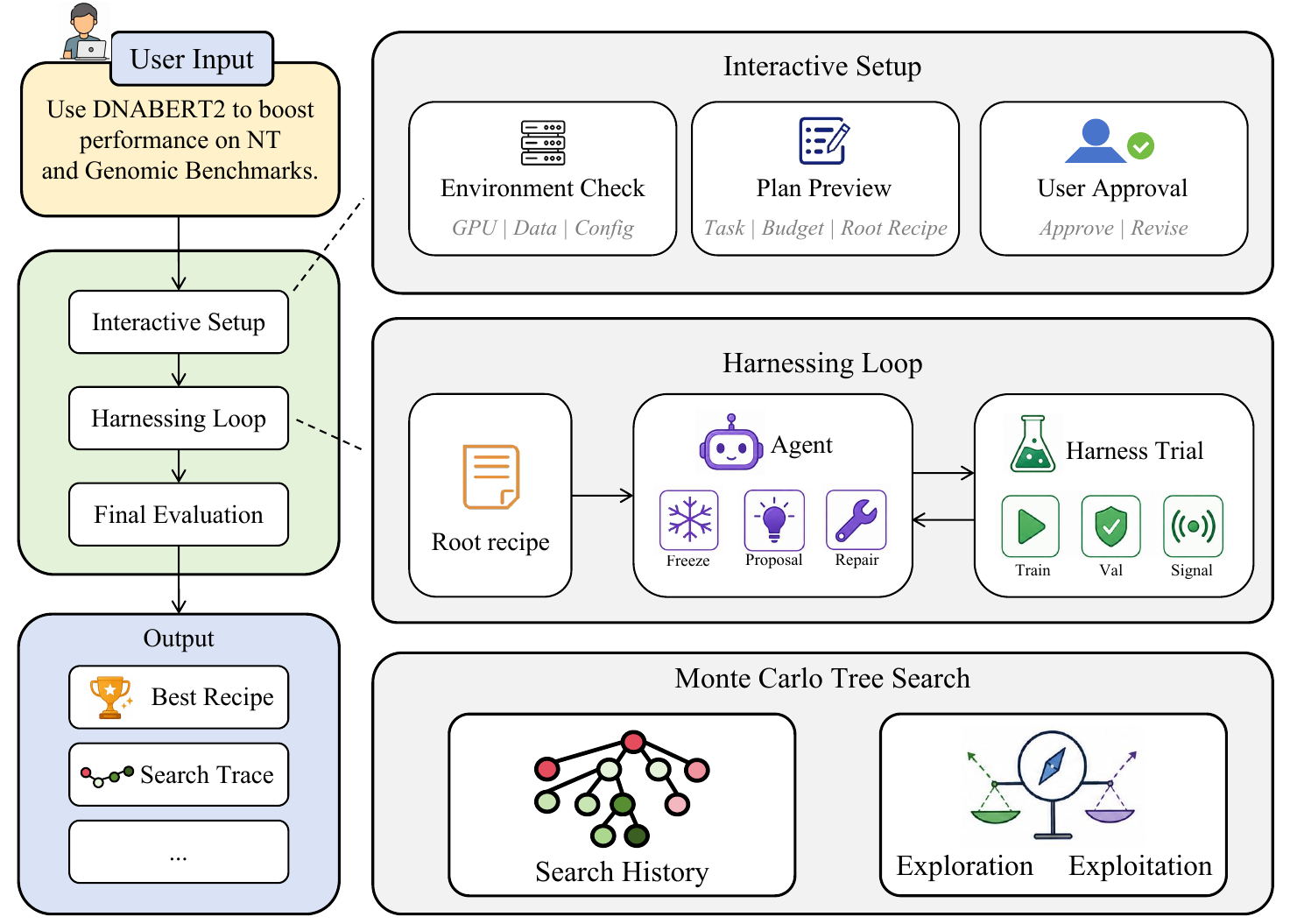}
    \caption{\centering Overview of the GenomeHarness framework.}
    \label{fig:overview}
\end{figure*}

Genome language models have become an increasingly important foundation for DNA sequence analysis \cite{benegas2025_glm_suvey1, consens2025transformers}. By pretraining on large-scale genomic sequences, models such as DNABERT \cite{ji2021dnabert}, DNABERT2 \cite{zhou2024dnabert}, Nucleotide Transformer \cite{dalla2025nt_model}, HyenaDNA \cite{nguyen2023hyenadna}, Caduceus \cite{schiff2024caduceus}, Enformer \cite{avsec2021enformer} and Evo \cite{nguyen2024evo1} provide reusable representations for downstream tasks including promoter and enhancer prediction, splice-site recognition, histone mark classification, and other regulatory sequence analyses. These models suggest a practical workflow in which researchers no longer need to train a sequence model from scratch, but can instead adapt a pretrained genome language model to a task of interest.

However, the practical use of genome language models is still bottlenecked by downstream adaptation. Most studies \cite{benegas2024gpnmsa, genalm} report a fixed fine-tuning recipe, specifying choices such as learning rate, scheduler, warmup, batch size, number of epochs, precision, and whether to use full fine-tuning or parameter-efficient tuning \cite{hu2021lora}. Such recipes are valuable starting points, but they are usually optimized for the models, datasets, and tasks considered in the original papers. For a new downstream application, it is often unclear whether a weak result reflects a limitation of the pretrained model itself or simply an under-optimized adaptation recipe. Manually searching for a better recipe can also be time-consuming and expensive, especially for downstream domain users who may not have the time, infrastructure, or machine-learning engineering bandwidth to repeatedly design, launch, debug, and compare fine-tuning runs.

This problem is especially pronounced in genome foundation model evaluation. Different benchmarks vary in dataset size, class balance, sequence length, metric, and biological objective \cite{dalla2025nt_model, grevsova2023genomic_benchmarks, marin2024bend_benchmark, patel2024darteval_benchmark, cheng2025dnalongbench}, and the best adaptation strategy may differ across both models and tasks. A recipe that works well for one splice-site task may not be optimal for histone mark prediction, and a setting tuned for one backbone may not transfer directly to another. Reliable use of genome language models therefore requires more than a single published configuration. It requires a systematic, budget-aware, and reproducible procedure that improves final performance, lowers the operational barrier of fine-tuning, and preserves a strict separation between validation-based recipe selection and final test evaluation.

Recent software frameworks have begun to lower the barrier to applying genome foundation models. gReLU \cite{lal2025grelu} provides a unified framework for DNA sequence modeling workflows, Genome-Factory \cite{wu2025genomefactory} integrates tuning, deployment, benchmarking, and interpretation for genomic foundation models, and S2F-agent \cite{li2026s2fagent} uses skill-grounded agent orchestration to translate open-ended genomics queries into reproducible sequence-to-function workflows. These systems improve accessibility and workflow integration, but they do not directly address systematic search for strong fine-tuning recipes under validation-only selection and test-isolated final evaluation.

At the same time, recent AI coding agents provide a promising interface for automating parts of this process, since they can propose recipe edits, inspect failures, and repair invalid or unstable runs \cite{chen2021codex, yao2023react, yang2024swe_agent}. Recent agentic systems further show that LLM-driven code generation becomes more effective when coupled with structured execution, evaluation, and feedback loops, as in AlphaEvolve \cite{novikov2025alphaevolve}, AutoHarness \cite{lou2026autoharness}, VCHarness \cite{cheng2026vcharness}, and automated AI research systems \cite{lu2026ai_autoresearch}. 

These observations point to a harnessed-agent approach for genome language model adaptation. The fine-tuning problem requires repeated recipe exploration and engineering support, while recent agentic harnesses show that LLM agents can be made more effective when their proposals are coupled to structured execution, evaluation, and feedback.

Here we present GenomeHarness, an agentic harness for reliable adaptation of genome language models, as summarized in Fig.~\ref{fig:overview}. GenomeHarness starts from literature-derived root recipes and converts downstream fine-tuning into a controlled search over executable training recipes. The AI agent proposes interpretable recipe edits, while the harness manages execution, state tracking, resource allocation, failure handling, metric collection, and final recipe freezing. Recipe search is organized by a Monte Carlo tree search (MCTS) \cite{browne2012survey_monte_carlo_tree_search} controller, which allocates trials across recipe lineages and progressively focuses computation on promising branches. The test set is accessed only after freezing, ensuring that final evaluation remains isolated from the search process.

We evaluated GenomeHarness across two pretrained genome language models, DNABERT2 and NTv2-100M-Multi, and two benchmark families, the NT Benchmark and Genomic Benchmarks \cite{grevsova2023genomic_benchmarks}. Across 52 model-task settings, GenomeHarness improves mean test MCC \cite{jurman2012MCC} in 47 settings, including 24 of 26 DNABERT2 settings and 23 of 26 NTv2-100M-Multi settings. The gains are especially pronounced on Genomic Benchmarks and on tasks where the literature-derived root recipe is unstable or poorly matched to the downstream data. In addition to final performance, we analyze search density, validation-test alignment, and trial-level search traces, showing that GenomeHarness can turn a limited compute budget into dense, auditable recipe evidence. In all, these results show that downstream adaptation is a central challenge for genome language models, and that harness-guided AI agents provide a practical route toward more systematic, reproducible, and accessible fine-tuning.

\section{Materials and Methods}

\subsection{Data Preparation}

% \footnote{NT Benchmark: https://huggingface.co/datasets/InstaDeepAI/nucleotide\_transformer\_downstream\_tasks\_revised} \footnote{Genomic Benchmarks: https://huggingface.co/katarinagresova/datasets}
We evaluated GenomeHarness on two benchmark families for genomic sequence classification: the NT Benchmark \cite{dalla2025nt_model} and Genomic Benchmarks \cite{grevsova2023genomic_benchmarks}. All datasets were downloaded from Hugging Face and processed using a unified split protocol. For each task, the original test set was kept unchanged and used only for final evaluation. The remaining training data were split into training and validation subsets with a 9:1 ratio. The validation set was used during the search stage for recipe selection and checkpoint selection, while the test set was never accessed during recipe search or freezing.

Genomic Benchmarks contain substantially larger datasets than the NT Benchmark, which can make repeated fine-tuning trials computationally expensive. To keep the search budget comparable across tasks, we capped the number of training candidates at 30,000 for each Genomic Benchmarks task before constructing the validation split. Specifically, if a task contained more than 30,000 training examples, we first sampled at most 30,000 examples according to the original label ratio, and then split the sampled examples into 27,000 training examples and 3,000 validation examples. The test sets were not downsampled. Detailed statistics for all tasks, including the numbers of training, validation, and test examples, label counts, and sequence length distributions, are summarized in Table~\ref{tab:data_stat}.

% Please add the following required packages to your document preamble:
% \usepackage{multirow}
% \usepackage{graphicx}
\begin{table*}[]
\centering
\resizebox{\textwidth}{!}{%
\begin{tabular}{cccccccc}
\hline
Name & Task & Train & Val & Test & Labels & Seq. Len Mean & Seq. Len Std \\ \hline
\multirow{18}{*}{NT Benchmark} & H2AFZ & 27000 & 3000 & 3000 & 2 & 1000 & 0 \\
 & H3K27ac & 27000 & 3000 & 1616 & 2 & 1000 & 0 \\
 & H3K27me3 & 27000 & 3000 & 3000 & 2 & 1000 & 0 \\
 & H3K36me3 & 27000 & 3000 & 3000 & 2 & 1000 & 0 \\
 & H3K4me1 & 27000 & 3000 & 3000 & 2 & 1000 & 0 \\
 & H3K4me2 & 27000 & 3000 & 2138 & 2 & 1000 & 0 \\
 & H3K4me3 & 15721 & 1747 & 776 & 2 & 1000 & 0 \\
 & H3K9ac & 20947 & 2327 & 1004 & 2 & 1000 & 0 \\
 & H3K9me3 & 24694 & 2744 & 850 & 2 & 1000 & 0 \\
 & H4K20me1 & 27000 & 3000 & 2270 & 2 & 1000 & 0 \\
 & enhancers & 27000 & 3000 & 3000 & 2 & 400 & 0 \\
 & enhancers\_types & 27000 & 3000 & 3000 & 3 & 400 & 0 \\
 & promoter\_all & 27000 & 3000 & 1584 & 2 & 300 & 0 \\
 & promoter\_no\_tata & 27000 & 3000 & 1372 & 2 & 300 & 0 \\
 & promoter\_tata & 4556 & 506 & 212 & 2 & 300 & 0 \\
 & splice\_sites\_acceptors & 27000 & 3000 & 3000 & 2 & 600 & 0 \\
 & splice\_sites\_all & 27000 & 3000 & 3000 & 3 & 600 & 0 \\
 & splice\_sites\_donors & 27000 & 3000 & 3000 & 2 & 600 & 0 \\ \hline
\multirow{8}{*}{Genomic Benchmarks} & demo\_coding\_vs\_intergenomic\_seqs & 27000 & 3000 & 25000 & 2 & 200 & 0 \\
 & demo\_human\_or\_worm & 27000 & 3000 & 25000 & 2 & 200 & 0 \\
 & dummy\_mouse\_enhancers\_ensembl & 872 & 96 & 242 & 2 & 2369 & 984.4 \\
 & human\_enhancers\_cohn & 18759 & 2084 & 6948 & 2 & 500 & 0 \\
 & human\_enhancers\_ensembl & 27000 & 3000 & 30970 & 2 & 267 & 122.9 \\
 & human\_ensembl\_regulatory & 27000 & 3000 & 57713 & 3 & 430 & 184.4 \\
 & human\_nontata\_promoters & 24387 & 2710 & 9034 & 2 & 251 & 0 \\
 & human\_ocr\_ensembl & 27000 & 3000 & 34952 & 2 & 326 & 107.9 \\ \hline
\end{tabular}%
}
\caption{Dataset statistics for the NT Benchmark and Genomic Benchmarks, including split sizes, label counts, and sequence length summaries.}
\label{tab:data_stat}
\end{table*}

\subsection{GenomeHarness Overview}

GenomeHarness is an agent-harness framework for reliable adaptation of genome language models. As illustrated in Fig.~\ref{fig:overview}, the system converts a natural-language user objective into a structured fine-tuning campaign, and organizes the campaign around three design principles: protocol locking, traceable recipe search, and validation-only selection before final test evaluation.

Before search begins, GenomeHarness performs an interactive setup that checks the required resources and artifacts, generates a reviewable campaign plan, and launches the campaign only after user approval. The approved plan fixes the pretrained backbone, benchmark definition, data split, label definitions, primary metric, search budget, and final evaluation policy. These components are treated as part of the experimental protocol rather than as tunable variables. The agent is therefore restricted to recipe-level decisions, while the harness enforces protocol constraints throughout the campaign.

GenomeHarness separates flexible agent decisions from deterministic execution. The agent proposes recipe edits, repairs failed trials, and freezes the selected recipe after search. The harness executes train-validation trials, records trial metadata, manages resources, tracks failures and repairs, and prevents test-set access before final evaluation. Recipe search is organized as a tree-structured history, where each evaluated recipe is linked to its parent recipe, edit description, validation signal, runtime status, and failure or repair record when applicable. 

The final output includes the frozen best recipe, final evaluation metrics, search trace and other campaign artifacts. A compact operational summary is provided in Supplementary Note S1, and a no-agent grid search ablation is provided in Supplementary Note S2.

\subsection{Controlled Recipe Search}

For each task, GenomeHarness starts from a literature-derived root recipe based on prior genome language model fine-tuning practices \cite{zhou2024dnabert,dalla2025nt_model,nguyen2023hyenadna,long2025mutbert,li2026omnidna}. The root recipe defines the default fine-tuning configuration and serves as the root node of the search tree. Its full detail is given in Supplementary Table S1. Candidate recipes are represented as traceable edits to either the root recipe or a previously evaluated recipe, rather than as independent configurations sampled from an unstructured space.

During search, the agent interacts with the harness through proposal, repair, and freeze actions. A proposal applies a concrete recipe edit to a selected parent recipe. A repair uses failure information from invalid, unstable, or failed trials to produce a constrained fix. A freeze action is used after search to select a completed recipe for final evaluation. In all cases, proposed changes must remain within the approved recipe-search scope and cannot alter the locked items.

Each candidate recipe is executed as a full training trial followed by validation. During search, all trials use search seed 42, and the raw validation MCC is used as the reward signal. The harness records the validation score, trial status, runtime information, recipe edit, parent-child relation, and failure or repair information. These records form the search history used by both the agent and the tree search controller in later iterations.

Recipe search is conducted under a fixed launch budget for each task. The budget limits the time window in which new trials can be launched, rather than terminating trials that are already running. GenomeHarness can execute concurrent multi-GPU trials during this window. Once the launch budget is reached, the harness stops launching new trials and waits for all running trials to finish before recipe freezing. This drain step ensures that the freeze decision uses all validation evidence generated within the approved search budget.

\subsection{MCTS-guided Search Controller}

GenomeHarness uses Monte Carlo tree search (MCTS) \cite{browne2012survey_monte_carlo_tree_search} to allocate search effort across recipe lineages. MCTS determines which evaluated recipe should be expanded next, the agent determines how to edit that recipe, and the harness determines whether the resulting candidate is valid and executable. This design differs from a flat hyperparameter grid because empirical feedback is reused through lineage-level statistics, and promising recipes can be progressively refined through traceable edits.

Each node $v$ in the search tree corresponds to a fine-tuning recipe $P_v$, and each edge corresponds to a recipe edit from a parent recipe to a child recipe. For a completed trial at node $v$, the reward is the validation MCC:
\begin{equation}
    r(v) = J_{\mathrm{val}}(P_v),
\end{equation}
where $J_{\mathrm{val}}(\cdot)$ denotes the validation metric computed by the harness. The test set is never used to compute search rewards.

For each node $v$, GenomeHarness maintains the visit count $N(v)$, cumulative reward $W(v)$, and empirical value
\begin{equation}
    Q(v) = \frac{W(v)}{N(v)}.
\end{equation}
At each search step, the controller traverses the tree using an upper confidence bound score. For a child node $u$ of a parent node $v$, the score is
\begin{equation}
    \mathrm{UCB}(u)
    =
    Q(u)
    +
    c
    \sqrt{
    \frac{\log\left(N(v)+1\right)}
    {N(u)+1}
    },
\end{equation}
where $c$ controls the exploration-exploitation trade-off. We set $c=0.05$ in our experiments.

Because agent-generated recipe edits form an open-ended action space, GenomeHarness uses progressive widening to control branching. Let $\mathcal{C}(v)$ denote the children of node $v$. A new child can be generated only when
\begin{equation}
    |\mathcal{C}(v)| < k \left(N(v)+1\right)^{\alpha}.
\end{equation}
Otherwise, the controller descends to an existing child according to the UCB score. We set $k=2$ and $\alpha=0.5$.

After a parent recipe is selected, the agent proposes a recipe patch conditioned on the parent and the accumulated search history:
\begin{equation}
    P_{\mathrm{child}}
    =
    P_{\mathrm{parent}}
    \oplus
    \Delta,
    \qquad
    \Delta \sim \pi_{\mathrm{agent}}(\cdot \mid P_{\mathrm{parent}}, \mathcal{H}_t),
\end{equation}
where $\Delta$ is the proposed edit, $\oplus$ denotes applying the edit, and $\mathcal{H}_t$ denotes the search history up to step $t$. The history includes evaluated recipes, validation scores, lineage relations, failure records, repair actions, and runtime metadata. Valid child recipes are launched as train-validation trials, and the observed reward is propagated back along the path to update visit counts and empirical values.

GenomeHarness also supports asynchronous execution. Since multiple trials may run concurrently, a recipe lineage can have pending descendants whose rewards are not yet available. To avoid repeatedly assigning concurrent trials to the same unfinished lineage, the controller applies a reservation penalty during parent selection:
\begin{equation}
    \mathrm{Score}(u)
    =
    \mathrm{UCB}(u)
    -
    \beta R(u),
\end{equation}
where $R(u)$ is the number of pending or running descendants under node $u$. We set $\beta=1.0$. This penalty encourages parallel trials to cover different recipe lineages while still allowing strong lineages to receive additional search effort after pending trials finish.

\subsection{Final Evaluation}

At the end of search, GenomeHarness freezes a single completed recipe based on validation performance. Recipe freezing marks the boundary between search and final evaluation. Final evaluation is performed only after recipe freezing. For each task, both the root recipe and the GenomeHarness-selected recipe are trained under the same final protocol with random seeds $\{42,43,44\}$. GenomeHarness reports the final test performance together with the frozen recipe, validation metrics, runtime information, failure and repair records, and the complete search trace documenting how the recipe was obtained.

\section{Results}

\subsection{Experimental Setup}

All GenomeHarness experiments were conducted on a local GPU server equipped with four NVIDIA L20 GPUs, each with 46 GB of memory. We used Codex powered by GPT-5.5 as the agent, with the intelligence setting configured to Extra High.
For each benchmark task, we allocated a fixed 1 hour search budget for recipe exploration. GenomeHarness could launch new search trials within the hour, while trials that had already started were allowed to finish. Final evaluation was performed separately after the best recipe had been frozen. During search, GenomeHarness used up to four GPUs to run concurrent single GPU trials under the approved campaign protocol. 

\subsection{GenomeHarness Improves Fine-tuning Performance}

% Please add the following required packages to your document preamble:
% \usepackage{multirow}
% \usepackage{graphicx}
\begin{table*}[]
\centering
\resizebox{\textwidth}{!}{%
\begin{tabular}{ccccccc}
\hline
\multirow{2}{*}{Task} & \multicolumn{3}{c}{DNABERT2} & \multicolumn{3}{c}{NTv2-100M-Multi} \\ \cline{2-7} 
 & Root & GenomeHarness & $\Delta_{\text{mean}}$ & Root & GenomeHarness & $\Delta_{\text{mean}}$ \\ \hline
H2AFZ & 0.4188 ± 0.1063 & 0.5165 ± 0.0033 & 0.0977 & 0.4955 ± 0.0165 & 0.4964 ± 0.0024 & 0.0009 \\
H3K27ac & 0.5135 ± 0.0037 & 0.5189 ± 0.0206 & 0.0055 & 0.4606 ± 0.0251 & 0.4981 ± 0.0244 & 0.0375 \\
H3K27me3 & 0.5901 ± 0.0223 & 0.6040 ± 0.0102 & 0.0139 & 0.5845 ± 0.0078 & 0.5983 ± 0.0149 & 0.0138 \\
H3K36me3 & 0.6265 ± 0.0157 & 0.6408 ± 0.0137 & 0.0144 & 0.5957 ± 0.0069 & 0.6300 ± 0.0060 & 0.0343 \\
H3K4me1 & 0.4870 ± 0.0147 & 0.5023 ± 0.0292 & 0.0154 & 0.4634 ± 0.0124 & 0.4939 ± 0.0088 & 0.0305 \\
H3K4me2 & 0.5679 ± 0.0029 & 0.5706 ± 0.0131 & 0.0028 & 0.5343 ± 0.0147 & 0.5634 ± 0.0091 & 0.0291 \\
H3K4me3 & 0.6098 ± 0.0115 & 0.6393 ± 0.0170 & 0.0294 & 0.6167 ± 0.0265 & 0.6542 ± 0.0071 & 0.0375 \\
H3K9ac & 0.4661 ± 0.1755 & 0.5618 ± 0.0339 & 0.0957 & 0.5252 ± 0.0262 & 0.5453 ± 0.0119 & 0.0200 \\
H3K9me3 & 0.4819 ± 0.0271 & 0.4795 ± 0.0147 & -0.0023 & 0.3381 ± 0.1143 & 0.4553 ± 0.0103 & 0.1172 \\
H4K20me1 & 0.6643 ± 0.0059 & 0.6634 ± 0.0159 & -0.0010 & 0.6339 ± 0.0128 & 0.6576 ± 0.0072 & 0.0237 \\
enhancers & 0.5418 ± 0.0071 & 0.5456 ± 0.0119 & 0.0038 & 0.5206 ± 0.0145 & 0.5145 ± 0.0017 & -0.0061 \\
enhancers\_types & 0.3039 ± 0.2611 & 0.5043 ± 0.0027 & 0.2005 & 0.4744 ± 0.0131 & 0.4895 ± 0.0137 & 0.0151 \\
promoter\_all & 0.7431 ± 0.0084 & 0.7641 ± 0.0068 & 0.0210 & 0.7563 ± 0.0118 & 0.7643 ± 0.0068 & 0.0081 \\
promoter\_no\_tata & 0.7334 ± 0.0328 & 0.7726 ± 0.0050 & 0.0392 & 0.7502 ± 0.0093 & 0.7618 ± 0.0063 & 0.0116 \\
promoter\_tata & 0.8401 ± 0.0184 & 0.8592 ± 0.0387 & 0.0191 & 0.8535 ± 0.0224 & 0.8569 ± 0.0234 & 0.0034 \\
splice\_sites\_acceptors & 0.8111 ± 0.0250 & 0.8376 ± 0.0106 & 0.0265 & 0.9651 ± 0.0047 & 0.9604 ± 0.0019 & -0.0047 \\
splice\_sites\_all & 0.7149 ± 0.2469 & 0.8549 ± 0.0031 & 0.1399 & 0.9627 ± 0.0017 & 0.9662 ± 0.0003 & 0.0035 \\
splice\_sites\_donors & 0.8426 ± 0.0225 & 0.8449 ± 0.0068 & 0.0024 & 0.9709 ± 0.0049 & 0.9680 ± 0.0013 & -0.0029 \\ \hline
Avg. MCC & 0.6087 ± 0.0331 & 0.6489 ± 0.0035 & 0.0402 & 0.6390 ± 0.0068 & 0.6597 ± 0.0017 & 0.0207 \\ \hline
\end{tabular}%
}
\caption{Final test performance on the NT Benchmark. For both the root recipe and the GenomeHarness-selected recipe, final evaluation was repeated with three random seeds, $\{42,43,44\}$. Each entry reports test MCC as mean $\pm$ standard deviation across seeds. $\Delta_{\mathrm{mean}}$ denotes the difference between the GenomeHarness mean MCC and the root mean MCC. Avg. MCC reports the average over all NT Benchmark tasks.}
\label{tab:res_nt}
\end{table*}

% Please add the following required packages to your document preamble:
% \usepackage{multirow}
% \usepackage{graphicx}
\begin{table*}[]
\centering
\resizebox{\textwidth}{!}{%
\begin{tabular}{ccccccc}
\hline
\multirow{2}{*}{Task} & \multicolumn{3}{c}{DNABERT2} & \multicolumn{3}{c}{NTv2-100M-Multi} \\ \cline{2-7} 
 & Root & GenomeHarness & $\Delta_{\text{mean}}$ & Root & GenomeHarness & $\Delta_{\text{mean}}$ \\ \hline
demo\_coding\_vs\_intergenomic\_seqs & 0.8519 ± 0.0066 & 0.8817 ± 0.0094 & 0.0299 & 0.8718 ± 0.0032 & 0.8877 ± 0.0017 & 0.0159 \\
demo\_human\_or\_worm & 0.9363 ± 0.0018 & 0.9488 ± 0.0012 & 0.0125 & 0.9155 ± 0.0039 & 0.9355 ± 0.0014 & 0.0200 \\
dummy\_mouse\_enhancers\_ensembl & 0.6671 ± 0.0190 & 0.7154 ± 0.0219 & 0.0483 & 0.6577 ± 0.0597 & 0.7273 ± 0.0082 & 0.0696 \\
human\_enhancers\_cohn & 0.5191 ± 0.0077 & 0.5242 ± 0.0101 & 0.0051 & 0.4625 ± 0.0089 & 0.4904 ± 0.0119 & 0.0279 \\
human\_enhancers\_ensembl & 0.6830 ± 0.0096 & 0.6853 ± 0.0150 & 0.0023 & 0.6587 ± 0.0137 & 0.6794 ± 0.0063 & 0.0206 \\
human\_ensembl\_regulatory & 0.5906 ± 0.1276 & 0.7262 ± 0.0572 & 0.1356 & 0.8843 ± 0.0037 & 0.8844 ± 0.0103 & 0.0001 \\
human\_nontata\_promoters & 0.7862 ± 0.0335 & 0.8968 ± 0.0060 & 0.1106 & 0.7939 ± 0.0091 & 0.8115 ± 0.0075 & 0.0176 \\
human\_ocr\_ensembl & 0.2504 ± 0.2313 & 0.5343 ± 0.0093 & 0.2839 & 0.4254 ± 0.0572 & 0.4682 ± 0.0107 & 0.0429 \\ \hline
Avg. MCC & 0.6606 ± 0.0296 & 0.7391 ± 0.0093 & 0.0785 & 0.7087 ± 0.0047 & 0.7355 ± 0.0028 & 0.0268 \\ \hline
\end{tabular}%
}
\caption{Final test performance on Genomic Benchmarks. For both the root recipe and the GenomeHarness-selected recipe, final evaluation was repeated with three random seeds, $\{42,43,44\}$. Each entry reports test MCC as mean $\pm$ standard deviation across seeds. $\Delta_{\mathrm{mean}}$ denotes the difference between the GenomeHarness mean MCC and the root mean MCC. Avg. MCC reports the average over all Genomic Benchmarks tasks.}
\label{tab:res_gb}
\end{table*}

Tables~\ref{tab:res_nt} and~\ref{tab:res_gb} summarize the final test MCC of the root recipe and the GenomeHarness-selected recipe on the NT Benchmark and Genomic Benchmarks, respectively. A validation-test alignment analysis is provided in Supplementary Note S3. Results are reported as mean $\pm$ standard deviation across final evaluation seeds, and the delta column reports the difference between the GenomeHarness mean MCC and the root mean MCC. Across 52 model-task settings, GenomeHarness improved the mean test MCC in 47 settings.

At the benchmark level, the improvement pattern was observed for both model backbones, although the magnitude varied across models and benchmark families.
For DNABERT2, GenomeHarness improved 24 out of 26 settings.
The average MCC increased from $0.6087$ to $0.6489$ on the NT Benchmark ($\Delta=+0.0402$), and from $0.6606$ to $0.7391$ on Genomic Benchmarks ($\Delta=+0.0785$).
For NTv2-100M-Multi, GenomeHarness improved 23 out of 26 settings.
The average MCC increased from $0.6390$ to $0.6597$ on the NT Benchmark ($\Delta=+0.0207$), and from $0.7087$ to $0.7355$ on Genomic Benchmarks ($\Delta=+0.0268$).
The larger gains for DNABERT2, especially on Genomic Benchmarks, suggest that recipe search is particularly useful when the default fine-tuning recipe is less well matched to the downstream task or model-specific optimization behavior.

The largest improvements occurred on tasks where the root recipe appeared poorly matched or sensitive to random seed.
The clearest example is human\_ocr\_ensembl, where GenomeHarness improved DNABERT2 from $0.2504 \pm 0.2313$ to $0.5343 \pm 0.0093$.
Similar stabilization was observed on DNABERT2 enhancers\_types, where MCC increased from $0.3039 \pm 0.2611$ to $0.5043 \pm 0.0027$, and on DNABERT2 splice\_sites\_all, where MCC increased from $0.7149 \pm 0.2469$ to $0.8549 \pm 0.0031$.
For NTv2-100M-Multi, the largest gain appeared on H3K9me3, where GenomeHarness improved the MCC from $0.3381 \pm 0.1143$ to $0.4553 \pm 0.0103$.
These cases suggest that a weak or unstable root result does not necessarily imply that the pretrained model lacks useful representations; it may instead reflect a brittle or under-optimized adaptation recipe.

The smallest positive changes were observed on tasks where the GenomeHarness-selected recipe was nearly tied with the root recipe or where the remaining improvement margin was limited.
For example, DNABERT2 showed small gains on human\_enhancers\_ensembl ($\Delta=+0.0023$), splice\_sites\_donors ($\Delta=+0.0024$), and H3K4me2 ($\Delta=+0.0028$).
For NTv2-100M-Multi, the changes were also minimal on human\_ensembl\_regulatory ($\Delta=+0.0001$), H2AFZ ($\Delta=+0.0009$), and promoter\_tata ($\Delta=+0.0034$).
These results indicate that GenomeHarness does not produce large gains uniformly across all tasks; rather, its benefit depends on how much optimization room the root recipe leaves.

The few non-improved settings were all on the NT Benchmark and were small in magnitude: DNABERT2 on H3K9me3 ($\Delta=-0.0023$) and H4K20me1 ($\Delta=-0.0010$), and NTv2-100M-Multi on enhancers ($\Delta=-0.0061$), splice\_sites\_acceptors ($\Delta=-0.0047$), and splice\_sites\_donors ($\Delta=-0.0029$).
For the two NTv2 splice-site tasks, the root recipe already achieved MCC values above $0.96$, so small validation-test fluctuations can dominate the observed difference under a fixed search budget.
Overall, the results show that GenomeHarness improves mean test performance in most settings, with the largest gains appearing when the root recipe is unstable, seed sensitive, or poorly matched to the downstream task.

\subsection{GenomeHarness Discovers Non-obvious Fine-tuning Recipes}

Beyond final test performance, we examined how GenomeHarness identifies the recipes selected for final evaluation.
A useful recipe search system should not only improve the validation objective, but also generate enough empirical evidence within a practical budget and make the selected recipe traceable to concrete search decisions.
We therefore analyzed both the search coverage produced by GenomeHarness and representative search trees leading to frozen recipes.

\begin{figure}[ht]
    \centering
    \includegraphics[width=0.8\linewidth]{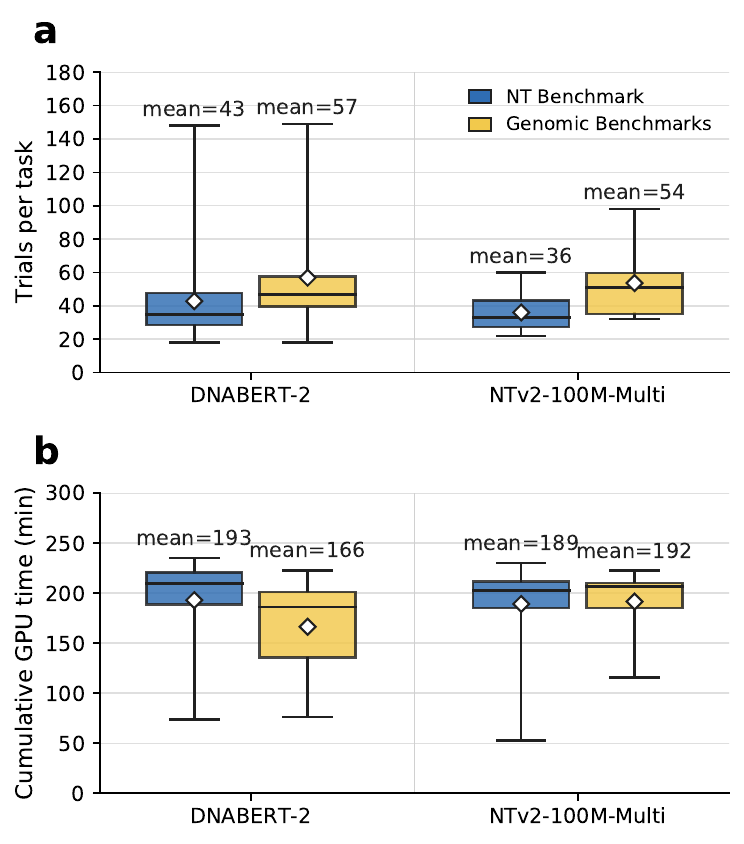}
    \caption{Search coverage under the fixed 1 hour launch budget.
    (a) Number of completed search trials per task.
    (b) Cumulative GPU time per task, summed across completed single-GPU train-validation trials.
    Boxes summarize the distribution across tasks within each model and benchmark family, and diamonds indicate group means.}
    \label{fig:search_efficiency}
\end{figure}

Figure~\ref{fig:search_efficiency} summarizes the search coverage under the fixed 1 hour launch budget.
Although new trial launches stopped after 1 hour, GenomeHarness could run multiple single-GPU trials in parallel and drain already-launched trials after the launch deadline.
As a result, it completed dozens of full train-validation recipe evaluations per task.
On average, GenomeHarness completed 43 trials for DNABERT2 on the NT Benchmark, 57 trials for DNABERT2 on Genomic Benchmarks, 36 trials for NTv2-100M-Multi on the NT Benchmark, and 54 trials for NTv2-100M-Multi on Genomic Benchmarks.
The corresponding cumulative GPU time reached 193, 166, 189, and 192 minutes per task, respectively.
Thus, the frozen recipe was selected from a dense set of completed and logged recipe evaluations rather than from a small number of manual adjustments.

\begin{figure*}[ht]
    \centering
    \includegraphics[width=\textwidth]{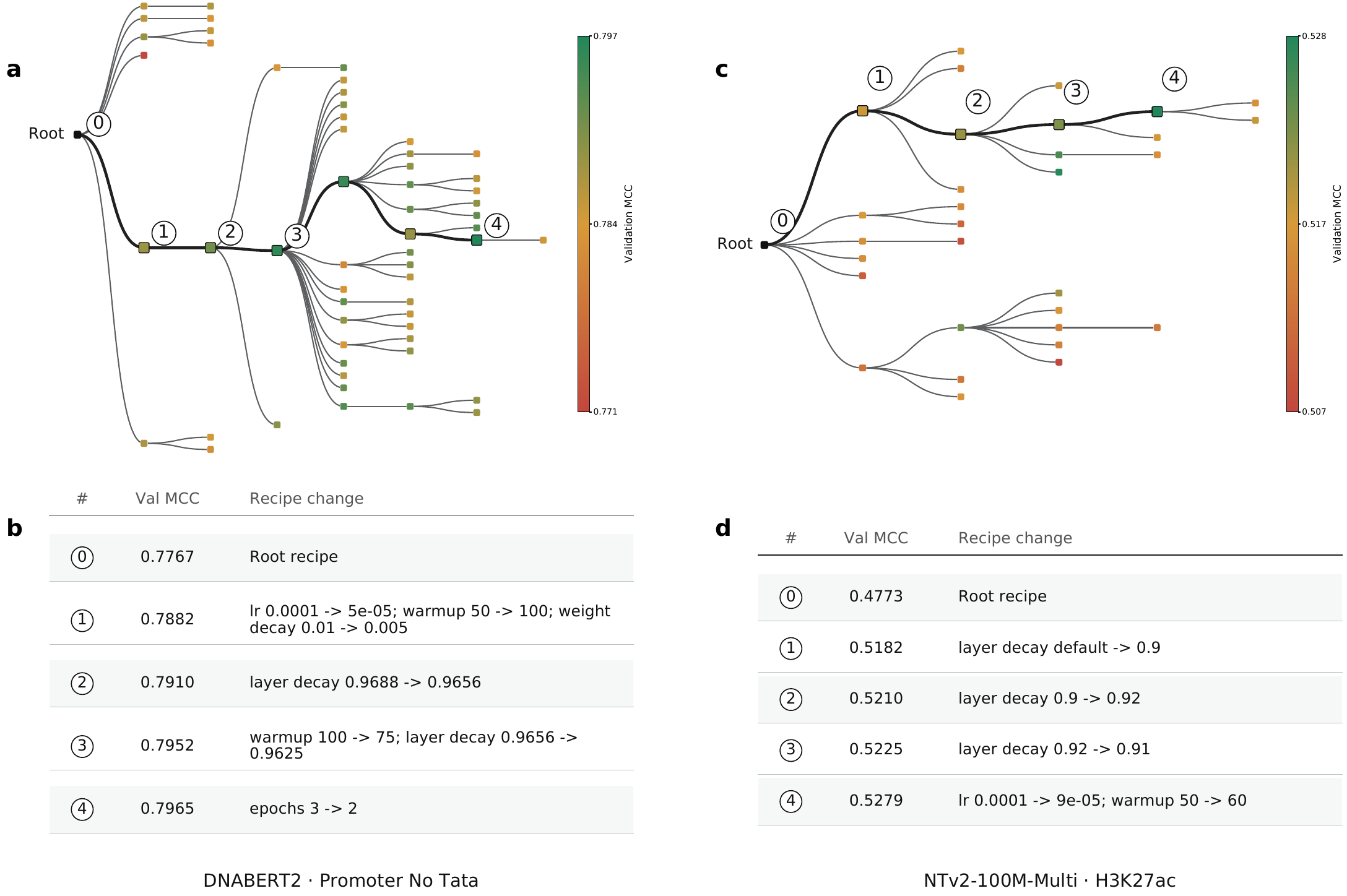}
    \caption{Representative GenomeHarness search trees and selected recipe lineages.
    (a), (b) Search tree and highlighted recipe lineage for DNABERT2 on promoter\_no\_tata.
    (c), (d) Search tree and highlighted recipe lineage for NTv2-100M-Multi on H3K27ac.
    In panels (a) and (c), each node denotes an evaluated candidate recipe and each edge denotes a recipe edit from a parent recipe to a child recipe.
    Node color indicates validation MCC, and the highlighted path marks the lineage leading to the frozen recipe.
    Panels (b) and (d) list the validation MCC and corresponding recipe changes along the highlighted lineage.}
    \label{fig:supp_search_tree}
\end{figure*}

To inspect the structure of recipe discovery, Fig.~\ref{fig:supp_search_tree} visualizes representative search trees together with the recipe changes along the lineage leading to the frozen recipe.
In each tree, nodes correspond to evaluated candidate recipes and edges correspond to recipe edits from a parent recipe to a child recipe.
Node color indicates validation MCC, and the highlighted path marks the selected lineage.

The two examples show that GenomeHarness does not treat candidate recipes as independent configurations.
For DNABERT2 on promoter\_no\_tata, the selected lineage first improves the root recipe by modifying learning rate, warmup, and weight decay, and then further refines layerwise decay, warmup, and training epochs.
For NTv2-100M-Multi on H3K27ac, the selected lineage mainly refines layerwise decay before making a smaller learning-rate and warmup adjustment.
These recipes are non-obvious in the practical sense that they are not single default settings or one-step manual changes, but multi-step, task-specific recipe lineages discovered through validation-guided search.
Together, the search coverage and lineage visualizations show that GenomeHarness converts a short wall-clock budget into dense, auditable, and traceable recipe evidence.

\subsection{GenomeHarness Search Traces Reveal Progressive Recipe Improvement}

\begin{figure*}[ht]
    \centering
    \includegraphics[width=\linewidth]{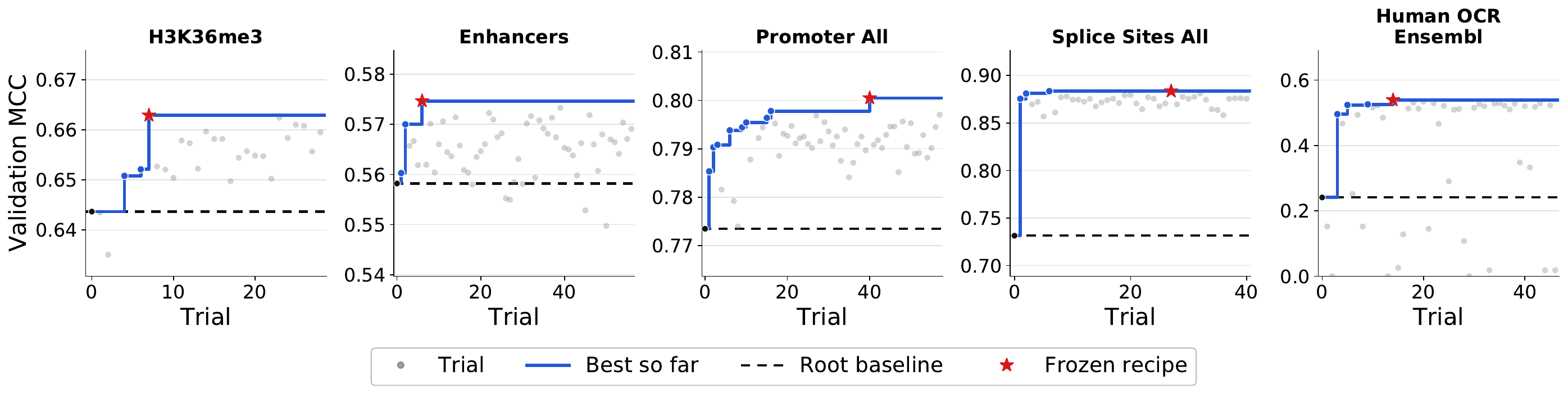}
    \caption{Representative GenomeHarness search traces on DNABERT2 tasks. Each panel shows the validation MCC of candidate recipes evaluated during the search stage. Gray points denote individual completed trials, the blue curve denotes the running best validation MCC, the dashed horizontal line denotes the root recipe baseline, and the red star marks the frozen recipe selected for final evaluation. The stepwise best-so-far curves illustrate how GenomeHarness accumulates trial-level validation feedback and identifies stronger recipes before held-out test evaluation.}
    \label{fig:search_dynamics}
\end{figure*}

We further inspected trial-level search traces to understand how GenomeHarness accumulates validation evidence during recipe search.
Figure~\ref{fig:search_dynamics} shows representative DNABERT2 search trajectories across five task types: a histone mark task (H3K36me3), an enhancer task (enhancers), a promoter task (promoter\_all), a splice-site task (splice\_sites\_all), and a Genomic Benchmarks task (human\_ocr\_ensembl).
% Because the best-so-far curve is updated only when a completed trial improves over the previous best, its stepwise increases directly indicate when GenomeHarness discovers stronger recipes during the search.

Across these tasks, GenomeHarness does not simply evaluate a fixed list of independent recipes.
Instead, the traces show an exploratory and feedback-driven refinement process.
Many evaluated recipes fall below the current best and remain visible as part of the search record, while stronger recipes update the best-so-far curve and become candidates for freezing.
On H3K36me3 and enhancers, GenomeHarness identifies improvements over the root baseline early in the search and then continues to evaluate additional candidates around the improved region.
On promoter\_all, the best-so-far curve improves through several smaller steps, illustrating gradual refinement rather than a single one-shot gain.
On splice\_sites\_all, the search rapidly moves far above the root baseline and then performs additional refinement before freezing.
Together, these trajectories show that GenomeHarness accumulates trial-level evidence and progressively selects stronger recipes from the explored candidate space.

The human\_ocr\_ensembl task provides a more difficult example.
The root recipe starts from a weak validation baseline, and many evaluated trials remain low-performing, indicating that this task contains substantial optimization risk under naive recipe choices.
Nevertheless, GenomeHarness identifies a substantially stronger recipe during the search, producing a clear upward jump in the best-so-far curve before freezing.
This behavior is consistent with the final test results in Table~\ref{tab:res_gb}, where human\_ocr\_ensembl shows one of the largest gains for DNABERT2.
This case highlights why systematic recipe exploration is useful for genomic fine-tuning: a single manually specified default recipe may fail badly on some tasks, whereas a harnessed search process can still recover a stronger configuration under the same validation-only protocol.

% Importantly, the search trajectories in Figure~\ref{fig:search_dynamics} are based only on validation MCC.
% The held-out test set is not used during search, repair, or recipe freezing. Thus, this figure provides process-level evidence for validation-guided recipe refinement, while the final test results in Tables~\ref{tab:res_nt} and~\ref{tab:res_gb} measure whether the selected frozen recipes generalize.
% The traces summarize trial order and best-so-far progress; a representative search tree is provided in Supplementary Note~S3 to show how evaluated recipes are organized into parent and child lineages.

% =========================================================
% Discussion
% =========================================================

\section{Discussion and Conclusion}

In this study, we introduced GenomeHarness, an agentic harness designed to improve the downstream adaptation of pretrained genome language models. Although these models provide reusable sequence representations, their practical use is still limited by the difficulty of turning a pretrained backbone into a reliable task-specific predictor. Downstream performance can depend strongly on fine-tuning recipes, and a weak result under a literature-derived default recipe may reflect a mismatched or unstable adaptation procedure rather than a limitation of the pretrained model itself. Beyond recipe choice, users still need to complete a series of practical steps, including preparing task data, setting up model-specific environments, managing compute resources, diagnosing failed or unstable runs, and comparing stochastic results without using the held-out test set. This process can be demanding for researchers who mainly focus on biological questions rather than machine-learning engineering.

GenomeHarness addresses this bottleneck by converting downstream adaptation from manual trial-and-error into a controlled search over executable fine-tuning recipes. Its central design is the coupling of three complementary components: MCTS allocates search effort across recipe lineages, the agent proposes concrete recipe edits and repairs conditioned on previous outcomes, and the harness executes trials, records evidence, manages resources, and enforces the approved experimental protocol. This design differs from a flat grid or random search because candidate recipes are organized as traceable refinements of earlier recipes rather than isolated configurations. It also differs from an unconstrained coding agent because each proposal must be validated through the same train-validation loop and becomes part of a structured search history.

Across two genome language models and two benchmark families, GenomeHarness improved mean test MCC in 47 of 52 model-task settings, including 24 of 26 DNABERT2 settings and 23 of 26 NTv2-100M-Multi settings. The gains were especially pronounced on Genomic Benchmarks and on tasks where the root recipe was weak or seed-sensitive, indicating that substantial downstream performance can remain hidden when a model is evaluated under only one default recipe. Importantly, these improvements were obtained under a fixed one-hour launch budget per task. Within this budget, GenomeHarness completed dozens of full train-validation recipe evaluations and selected the final recipe from dense, logged search evidence rather than from a small number of manual attempts.

The search traces further show that GenomeHarness improves adaptation through progressive recipe refinement. The selected recipes were often not single-step modifications, but multi-step lineages involving changes to learning rate, warmup, weight decay, layerwise decay, or training duration. This behavior is important because effective fine-tuning recipes are often task- and backbone-specific, and may emerge only after several empirical adjustments. Difficult cases such as human ocr ensembl illustrate this point clearly: many candidate recipes remained weak, but the harnessed search process still recovered a substantially stronger configuration. Conversely, the few non-improved settings were small in magnitude and mostly occurred when the root recipe was already strong or close to saturation, suggesting that GenomeHarness is most useful when the default recipe leaves meaningful optimization room.

Several limitations remain. First, the current search is bounded by the recipe axes, agent proposals, and compute budget available within a single campaign. Future systems could improve sample efficiency by reusing successful recipe motifs across related tasks or learning proposal policies from previous campaigns. Second, the current MCTS controller uses validation reward to guide search, but it does not yet explicitly model which recipe axes transfer across tasks, backbones, or benchmark families. More hierarchical or cost-aware search strategies could first explore broad recipe families and then refine promising axes more efficiently. Third, this study focuses on sequence classification tasks using two pretrained backbones, so broader validation is needed on longer-sequence tasks, variant effect prediction, regression objectives, and multimodal genomic settings. 

Overall, GenomeHarness shows that reliable use of genome language models requires not only strong pretrained backbones, but also systematic adaptation procedures. By combining MCTS-guided search, agent-generated recipe edits, and harness-controlled execution, GenomeHarness turns downstream fine-tuning into a traceable empirical search process and lowers the practical barrier to applying pretrained genome models across diverse biological tasks.

\section{Author contributions}
W.L. and Y.Z. conceived the study. W.L., Y.H., H.S. and J.F. performed analysis. Y.Z. supervised the project. W.L. and Y.Z. wrote the article. All authors read and approved the final article.

\section*{Acknowledgements}
This work is supported by the National Natural Science Foundation of China (No. 32500550), the Guangzhou-HKUST(GZ) Joint
Funding Program(2025A03I3865) and a Guangdong Provincial Project (2024QN11N085).

\section*{Competing interests}
The authors declare no competing interests.

\section*{Data and code availability}

The datasets and codebase used in this work can be accessed
at https://github.com/ai4nucleome/GenomeHarness.

\bibliographystyle{unsrtnat} 
\bibliography{reference}

\end{document}